\documentclass[showpacs,prb,twocolumn,floats,superscriptaddress]{revtex4}
\usepackage{graphicx}
\usepackage{amsmath}
\usepackage{epstopdf}
\usepackage{amssymb}
\usepackage{color}
\usepackage{ifpdf}

\newcommand{\eps}{\varepsilon}

\renewcommand{\i}{\mathrm{i}}

\begin{document}
\title{Bilayer graphene Hall bar with a pn-junction}

\author{S. P. Milovanovi\'{c}}\email{slavisa.milovanovic@gmail.com}
\affiliation{Departement Fysica, Universiteit Antwerpen \\
Groenenborgerlaan 171, B-2020 Antwerpen, Belgium}

\author{M.~Ramezani Masir}\email{mrmphys@gmail.com}
\affiliation{Departement Fysica, Universiteit Antwerpen \\
Groenenborgerlaan 171, B-2020 Antwerpen, Belgium}

\author{F.~M.~Peeters}\email{francois.peeters@uantwerpen.be}
\affiliation{Departement Fysica, Universiteit Antwerpen \\
Groenenborgerlaan 171, B-2020 Antwerpen, Belgium}
\begin{abstract}
We investigate the magnetic field dependence of the Hall and the bend resistances for a ballistic Hall bar structure containing a pn-junction sculptured from a bilayer of graphene. The electric response is obtained using the billiard model and we investigate the cases of bilayer graphene with and without a band gap. Two
different conduction regimes are possible: $i$) both sides of the junction have the same carrier type, and $ii$) one side of the junction is n-type while the other one is p-type. The first case shows Hall plateau-like features in the Hall resistance that fade away as the band gap opens. The second case exhibits a bend resistance that is asymmetric in magnetic field as a consequence of snake states along the pn-interface, where the maximum is shifted away from
zero magnetic field.
\end{abstract}


\date{\today}

\maketitle

\twocolumngrid

\section{Introduction}
The study of single-layer, bilayer, and multilayer graphene has been
intensified due to their drastically different electronic
properties from those of conventional semiconductors. Graphene
has a linear spectrum near the $K$ and $K'$ points \cite{f1,f2}
which cause perfect transmission through arbitrarily high and
wide barriers for normal incidence, referred to as Klein tunneling
\cite{f3,f4,f5,f7}. Another consequence is that single layer and bilayer graphene display an unconventional quantum Hall
effect\cite{g1,g2}. In contrast to carriers in single-layer
graphene, those in bilayer graphene possess a {\it quadratic}
spectrum near the K-points and show no Klein tunneling\cite{f3}.
Adsorbates and/or gate potentials \cite{g3,g4,g5} induce an energy
gap due to the tunnel coupling between the layers that is beneficial  for certain applications, e. g., for improving the
on/off ratio in carbon-based transistors. The metamaterial character
of pn-structures in graphene\cite{chei07} was pointed out, and
focusing of electronic waves was proposed\cite{mog10,has10}.
Moreover, snake states along the pn-interface were predicted analytically \cite{Ma1,Ma2} and experiments on such
systems were undertaken recently\cite{Marcus1,Marcus0}. The
metamaterial properties of the above mentioned pn-structures
resulted in the expectancy of controlling the electron wave
function, in particular, the width of electron beams by means of a
superlattice  known as collimation \cite{Cheol}. Qualitatively, the
metamaterial properties of pn-junctions in graphene can be
understood by inspecting classical trajectories\cite{A2}, or using
ray optics as it is called for the case of electromagnetic
phenomena\cite{A1}. Classical transport simulations were recently
\cite{A3, cMLG} presented for a single layer graphene Hall bar. Bilayer graphene exhibits a quadratic spectrum for a low energy which results in a very different transmission probability for potential barriers, i.e. absence of Klein tunneling. This motivated us to investigate the response of a bilayer graphene Hall
bar containing a pn-junction which has quantitative different transmission properties. The present results will be contrasted with those
of single layer graphene.

The paper is organized as follows. In Section \ref{Mod} our method
of solving the electronic transport of a Hall bar is introduced as well as the
procedure for obtaining the transmission and reflection
coefficients. In Section \ref{res} we present our numerical results
for the Hall and the bend resistances and analyze their behavior for
the case of bilayer graphene with and without a band gap in
its spectrum. Conclusions and remarks are given in Section
\ref{conc}.
\section{Model}
\label{Mod}
\begin{figure}[ht]
\begin{center}
\includegraphics[width=8.5cm]{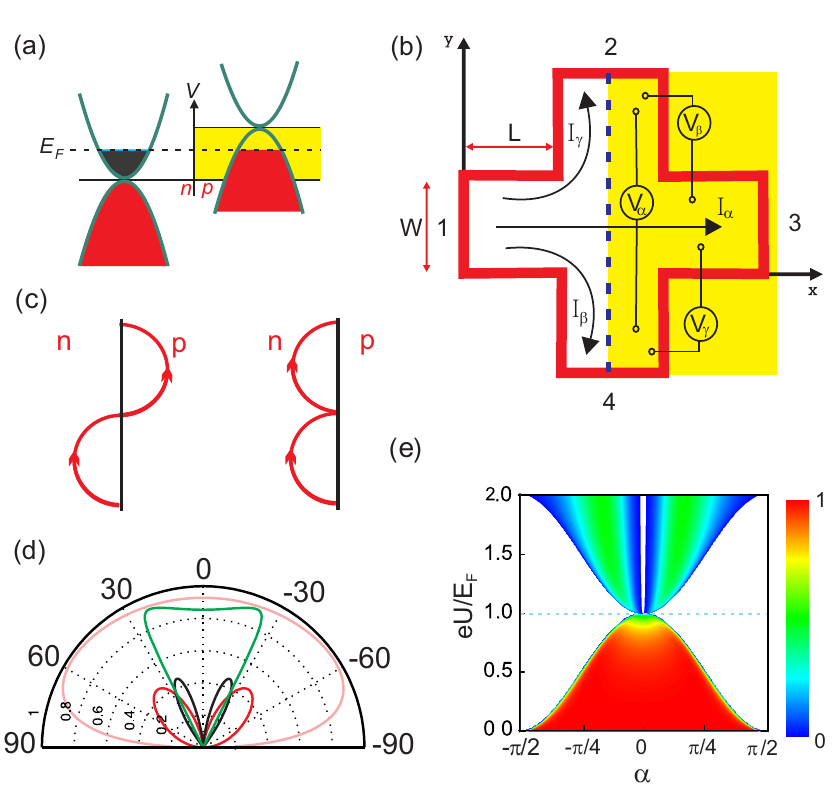}
\caption{(Color online) (a) Schematic model of a potential step
profile with the charge carrier bands and (b) the Hall bar model in the presence of this step potential. (c) Electron scattering on the pn-interface can result in snake orbits (left) and/or skipping orbits (right). (d) Angular dependence of the transmission for $eU/E_F=$ $0.2$ (pink), $0.8$ (green), $1.2$ (black), and $2.0$ (red). (e) Angular dependence of the transmission probability.}
\label{fig1}
\end{center}
\end{figure}

The system we are investigating is a $4$-terminal bilayer graphene
structure in the shape of a Hall bar, shown in Fig. \ref{fig1}(b).
The following typical parameters are used, $E_{F} = 50 meV$ is the Fermi energy, $v_{F} = 10^{6}$m/s is the Fermi velocity, and $W = L = 1 \mu$m, resulting in $\displaystyle{B_{0} = \frac{|E_{F}|}{e v_{F} W} =  0.05}$T,
 $\displaystyle{R_{0} =\frac{h}{4 e^{2}}\frac{\hbar v_{F}}{|E_{F}|W} = 0.117\frac{h}{4e^{2}}}$ which are taken as units for, respectively, the magnetic field and the resistance. The change in resistances when the same potential is
applied to both layers of bilayer graphene is examined, as well
as the situation when the band gap is opened as a consequence of the
different potentials applied to the top and the bottom layer of bilayer graphene. In order to obtain the transport properties of the bilayer
graphene  Hall bar structure in the presence of an external magnetic field we
implemented a semiclassical approach \cite{cBM,cSCA}. Electrons are assumed to
move ballistically and the billiard model is used to simulate their
motion. The trembling motion of electrons (Zitterbewegung) is
neglected due to its transient character which makes it observable
only on a femtosecond scale\cite{zit}.

The Hamiltonian describing the electron motion in bilayer graphene is given by
\begin{equation}\label{4BH}
H = \left(\begin{array}{c c c c}
U' & v(p_x+\i p_y) & t_{\perp} & 0 \\
v(p_x-\i p_y) & U' & 0 & 0 \\
t_{\perp} & 0 & U'' & v(p_x-\i p_y) \\
0 & 0 & v(p_x+\i p_y) & U'' \\
\end{array}\right) ,
\end{equation}
where $U'$ and $U''$ represent the applied potentials on the top and the bottom layer of the bilayer graphene structure, respectively,
$t_\perp\approx 0.4 eV$ is the interlayer hopping parameter and $v_{F}$ is
the Fermi velocity. 

\begin{figure}[htbp]
\begin{center}
\includegraphics[width=6cm]{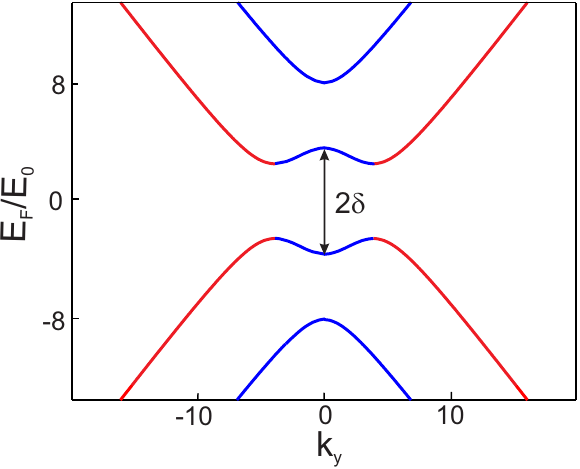}
\caption{(Color online) Energy dispersion relation for bilayer
graphene with an open gap $\delta$. Spectrum consists of two different wave
vectors: $k_+$ (red line) and $k_-$ (blue line). Plot
is made for $E_0 = 50meV$.} \label{fig_spb}
\end{center}
\end{figure}

Solving $H\Psi = E\Psi$ with $\Psi =
[\phi_a,~\phi_b,~\phi_c,~\phi_d]^T$ in both regions of the structure we derive the expression for the wave
vector in the $x$-direction as:
\begin{equation}\label{eq1}
k_{\sigma \pm} = \displaystyle{\left[ \eps_\sigma^2+\delta_\sigma^2 \pm \sqrt{4\eps_\sigma^2\delta_\sigma^2 + (\eps_\sigma^2-\delta_\sigma^2)t^2}-q^2 \right]^{1/2}},
\end{equation}
with $U_\sigma = (U'_\sigma+U''_\sigma)/(2\hbar v_F)$, $2\delta_\sigma=e(U'_\sigma-U''_\sigma)/(\hbar v_F)$, $\eps_\sigma = (E_F-eU_\sigma)/(\hbar v_F)$, $\sigma = \left\lbrace1,2\right\rbrace$ corresponds to the first or the second region, and $q$ is the
wave vector in the $y$-direction. If $U'_\sigma=U''_\sigma$ there is no gap in the
bilayer spectrum and $k_{\sigma\pm}$ transforms to a much simpler expression
\begin{equation}\label{eq1ng}
k_{\sigma \pm} = \displaystyle{\eps^2 \pm \left|\eps t\right|-q^2}.
\end{equation}
A plot of the energy bands for bilayer graphene when a gap is opened is shown in Fig.
\ref{fig_spb}. We see that the states around the band extremum
correspond to both $k_{\sigma+}$ and $k_{\sigma-}$ states.

Using the obtained wave vectors we can write the wave functions as
\begin{equation}\label{WF1}
\Psi_{1}^{\pm} = \Psi_R^\pm + r^\pm_+\Psi_L^+ +r^\pm_-\Psi_L^-,
\end{equation}
in the first region and
\begin{equation}\label{WF2}
\Psi_{2}^{\pm} = t^\pm_+\Psi_R^+ + t^\pm_-\Psi_R^- ,
\end{equation}
in the second region, where
\begin{equation}\label{WF3}
\Psi_R^\pm = N_{\sigma\pm}
\left(
\begin{array}{c}
\displaystyle{\frac{1}{t\xi_\sigma}\left\lbrace \xi_\sigma^2 -k_{F_{_\sigma \pm}}^2\right\rbrace}\\
\displaystyle{ \frac{1}{t(\eps_\sigma^2-\delta_\sigma^2)} \left[ k_{\sigma \pm} - \i q\right]\left\lbrace \xi_\sigma^2 -k_{F_{\sigma\pm}}^2\right\rbrace}\\
\displaystyle{1}\\
\displaystyle{\frac{1}{\xi_\sigma} \left[  k_{\sigma\pm} + \i q\right] }\\
\end{array}\right)e^{ik_{\sigma\pm} x}
\end{equation}
and
\begin{equation}\label{WF4}
\Psi_L^\pm = N_{_\sigma\pm}
\left(
\begin{array}{c}
\displaystyle{\frac{1}{t\xi_\sigma}\left\lbrace \xi_\sigma^2 -k_{F_{\sigma\pm}}^2\right\rbrace}\\
\displaystyle{ -\frac{1 }{t(\eps_\sigma^2-\delta_\sigma^2)} \left[ k_{\sigma\pm} + \i q\right]\left\lbrace \xi_\sigma^2 -k_{F_{\sigma\pm}}^2\right\rbrace}\\
\displaystyle{1}\\
\displaystyle{-\frac{1}{\xi_\sigma} \left[  k_{\sigma\pm} - \i q\right] }\\
\end{array}\right)e^{-ik_{\sigma\pm} x}
\end{equation}
with $\xi_\sigma = (\eps_\sigma-\delta_\sigma)$ and $N_{\sigma\pm}$ a normalization constant given by
\begin{equation}
\displaystyle{N_{\sigma\pm}= \left[evW\frac{2k_{\sigma\pm}(\xi_\sigma^2-k_{F_{\sigma\pm}}^2)^2+2k_{\sigma\pm} t^2(\eps_\sigma^2-\delta_\sigma^2)}{t^2(\eps_\sigma^2-\delta_\sigma^2)\xi_\sigma}\right]^{-1/2}},
\end{equation}
and $k_{F_{\sigma\pm}} = \sqrt{k_{\sigma\pm}^2+q^2}$. Equating the wave functions at the
boundary $x= 0$ we obtain the reflection and transmission
coefficients $r_\pm^\pm$ and $t_\pm^\pm$. Total reflection and
transmission coefficients are, respectively, given by
\begin{equation}\label{rec}
\displaystyle{R = \frac{1}{2}Tr(\mathbf{rr^\dagger})}
\end{equation}
and
\begin{equation}\label{trc}
\displaystyle{T =\frac{1}{2} Tr(\mathbf{tt^\dagger}).}
\end{equation}
Figs. \ref{fig1}(c) and (d) show transmission probability for
bilayer graphene without a gap. Plots are in agreement with the
results presented in Ref. \onlinecite{cHET}. Figures show that an
electron with energy lower than the height of the potential step
will be fully reflected from the step potential for the case of
normal incidence. Note that for single layer graphene, due to Klein
tunneling, this probability was 1.

\begin{figure}[htbp]
\begin{center}
\includegraphics[width=7.5cm]{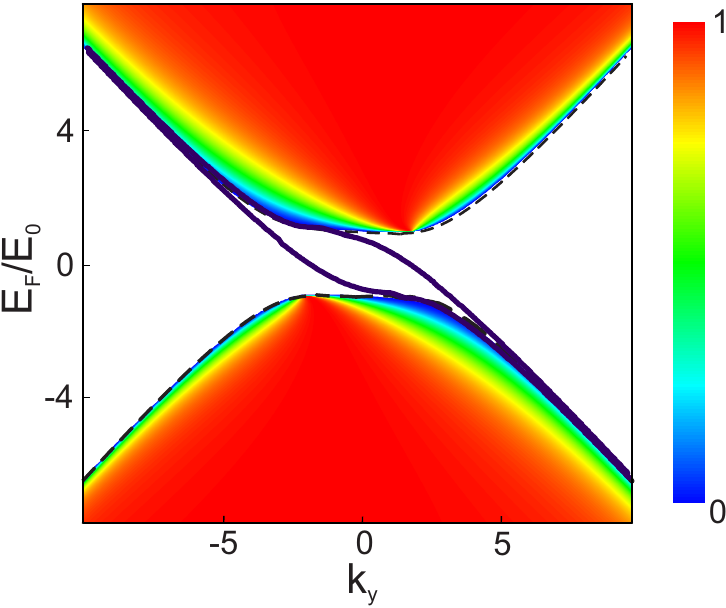}
\caption{(Color online) Transmission probability for bilayer graphene with $U_1=U_2 = 0$, $E_0 = 50meV$ and $\delta_1 =-\delta_2$. Purple curves present the chiral states.} \label{chrl}
\end{center}
\end{figure}

Fig. \ref{chrl} shows the transmission probability versus the Fermi energy $E_F$ and wave vector $\mathbf{k}$ for bilayer graphene with band gap in both regions $\delta_1 = -\delta_2$. Compared with the transmission obtained in the case of a gapless system, shown in Fig. \ref{fig1}(e), significant differences can be observed. The most important one is the absence of symmetric behavior of transmission probability with respect to the sign change of $k_y$ together with the appearance of chiral states.\cite{cCHRL1,cCHRL2} It is shown that these localized states appear at the boundary, with energies inside the energy gap corresponding to unidirectional motion of electrons.

We inject a large number of electrons (typically $10^5$) from each terminal of our $4$-terminal structure with initial velocity
$v_F$ with random position and random angle with a weighted angular distribution $P(\alpha) = 1/2\cos(\alpha)$. The motion of ballistic particles
is determined by the classical Newton equation of motion, which is
justified for the case $l_\phi<W<l_e$ where $l_\phi$ is the phase
coherence length and $l_e$ the mean free path. The electron mean free path can be
calculated as $l_e=(\hbar /e)\mu (\pi n_s)^{1/2}> 1\mu m$, with
$\mu$ the mobility which is about $ 200,000$
$cm^2V^{-1}s^{-1}$ at a carrier density $n_s=10^{12} cm^{-2}$.
The transmission of electrons through the potential step is
calculated quantum mechanically using the Dirac Hamiltonian. In our
simulation when a particle hits the potential step we calculate the
transmission probability using Eq. \eqref{trc}. If the particle is reflected back  mirror reflection is assumed. This means that reflected angle, $\alpha_r$, is related to the incident one as $\alpha_r = -\alpha_i$. Similar specular reflection is assumed at the borders of the structure. If, on the other
hand, the particle passes through the step the transmission angle is calculated using energy and momentum conservation.

To investigate what happens if we sculpture a Hall bar from bilayer
graphene our model needs to include all its unique properties.
Compared with single layer graphene this material shows several
essential differences:

$i$) In our simulation we will solve the classical equations of
motion but to do so proper substitution for a mass term needs to be found.
This is done using the Hamiltonian of our system.
 The difference with the single layer graphene case comes
from the fact that the dispersion relation for Dirac electrons is
different, now we have $v_F^2p_\sigma^2/t_{\perp} = (E_F-U_\sigma)$ (while in the case of single layer
graphene one has $v_F p_\sigma = (E_F-U_\sigma)$). Using this equation we can express the effective mass $m$ as $m =  t_{\perp}/(2v_F^2)$,
and insert it into the equations of motion.

$ii$) In the case of monolayer graphene we have a linear dependence
between energy and momentum which leads to a density of
states (DOS) that is linear in energy. In the case of bilayer
graphene the dependence between energy and momentum is quadratic and
the DOS is energy independent, $D(E) = A/(4\pi v_F^2\hbar^2 ) $, where $A$ is the surface area.

$iii$) Transmitted angle $\alpha_{t}$ is related with the incident angle $\alpha_{i}$ as,
\begin{equation}\label{Angle}
    \sqrt{|E-U_{i}|}\sin{\alpha_{i}} =
    \sqrt{|E-U_{t}|}\sin{\alpha_{t}},
\end{equation}
which is nothing else then Snell's law.

$iv$) Movement of particles after
transmission through the potential step shows another difference between monolayer and bilayer graphene. In single layer graphene the particle that transmits through the potential step will have a new angle -$\alpha_t$ if $s_i \neq s_t$, where $s_{i,t} = sgn(E_F-eU_{i,t})$ due to its metamaterial properties. Bilayer graphene is not a metamaterial and therefore this will no longer be
the case.

Transport properties of the system are calculated using the Landauer-B\"{u}ttiker formalism. For this purpose we need to find the electron transmission probabilities between the different leads of the Hall bar structure.
The probability that an electron injected from terminal $j$ will end up in
terminal $i$ is given by $T_{ij}$. These transmission probabilities are then
used in the Landauer-B\"{u}ttiker formula in order to calculate the current in
terminal $i$,
\begin{equation}
 I_i = \frac{e}{h}\left[(N_i-T_{ii})\mu_i-\sum_{j\neq i}T_{ij}\mu_j\right],
\end{equation}
 here $N_i$ is the number of occupied transport channels, which depends on $E_F$,
 $\mu_i$ and $\mu_j$ are the chemical potentials of the reservoirs $i$ and $j$,
 respectively,  $e$ is the electron charge and $h$ is Planck's constant.
 Eliminating the chemical potentials we can derive expressions for the different resistances,
 \begin{equation}
 R_{mn,kl} = \frac{h}{e^2}\frac{T_{km}T_{ln}-T_{kn}T_{lm}}{D},
 \label{eq2}
 \end{equation}
with $D =\alpha_{11}\alpha_{22}-\alpha_{12}\alpha_{21}$ and
\begin{equation}
\begin{array}{l}
\alpha_{11} = \left[(N_i-T_{11})S-(T_{14}+T_{12})(T_{41}+T_{21})\right]/S \\
 \alpha_{12} = (T_{12}T_{34}-T_{14}T_{32})/S \\
   \alpha_{21} =(T_{21}T_{43}-T_{41}T_{23})/S \\
 \alpha_{22} = \left[(N_i-T_{22})S -(T_{21}+T_{23})(T_{32}+T_{21})\right]/S,
\end{array}
\end{equation}
where $ S = T_{12}+T_{14}+T_{32}+T_{34}$.
\section{Results and discussion}
\label{res}

We are interested to obtain the Hall resistance, $R_H = R_\alpha = R_{13,24}$, the bend resistances $R_B = R_\beta =R_{14,32}$ and
$R_G = R_\gamma = R_{12,43}$, and their counterparts obtained by switching the voltage and current probes: $R_{HH} = R_{24,13}$, $R_{BB} = R_{32,14}$, and $R_{GG} = R_{43,12}$. These resistances are defined by $R_i = V_i/I_i$ ($i=\alpha,\beta,\gamma$), as shown in Fig. \ref{fig1}(b). For simplicity reasons, we consider a potential step where in  the first region we take $U_{1}=0$ and $\delta_1=0$ while the values of the potential in the second region $U = U_2$ and $\delta = \delta_2$ will be varied.

Although the four-fold symmetry of the system is broken by the presence of the pn-interface still some symmetry relations can be found between the different resistances. From Fig. \ref{fig1}(b) we see that: $T_{j2}(B)=T_{j4}(-B)$ and $T_{j2}(B)=T_{4j}(-B)$ with $j = \lbrace 1,3 \rbrace$, as well as $T_{24}(B)=T_{42}(-B)$, $T_{31}(B)=T_{31}(-B)$ and $T_{13}(B)=T_{13}(-B)$. Inserting these equalities into the expressions for $R_B$ and $R_G$ we derive
\begin{equation}
\begin{array}{l}
 D R_{14,32}(B) = T_{31}(B)T_{24}(B)-T_{34}(B)T_{21}(B)= \\
 T_{31}(-B)T_{42}(-B)-T_{32}(-B)T_{41}(B) = -R_{12,43}(-B)D.
 \end{array}
 \label{rbprop}
 \end{equation}
 In a similar way we obtain $R_{GG}(B)= -R_{BB}(-B)$. Therefore, it suffices to investigate only the behavior of $R_B$ and $R_{BB}$ while the behavior of the other bend resistances can be obtain using these simple symmetry transformations. 
\begin{figure*}[htbp]
\begin{center}
\includegraphics[width=16cm]{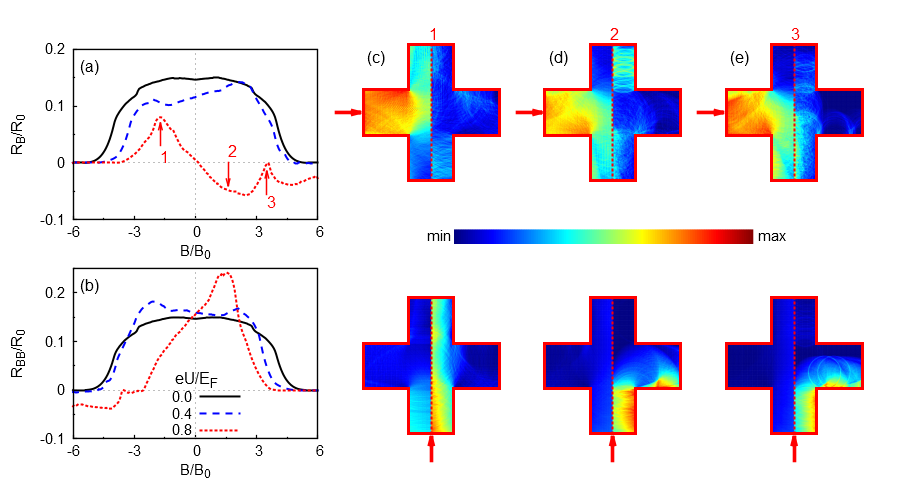}
\caption{(Color online) The bend resistances (a) $R_B$ and (b) $R_{BB}$ for different values of applied potential $U$ given in the inset of (b). (c)-(e) Electron current density plots for magnetic field values indicated in (a). Arrows indicate the injection lead.} \label{fig4}
\end{center}
\end{figure*}
\begin{figure*}[htbp]
\begin{center}
\includegraphics[width=16cm]{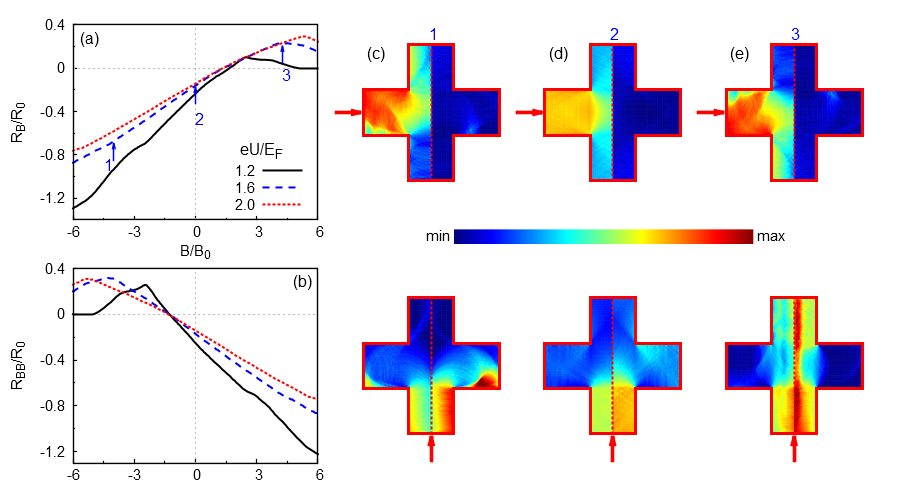}
\caption{(Color online) The same as Fig. \ref{fig4} but now for $eU>E_F$.} \label{fig4b}
\end{center}
\end{figure*}
\begin{figure*}[htbp]
\begin{center}
\includegraphics[width=14cm]{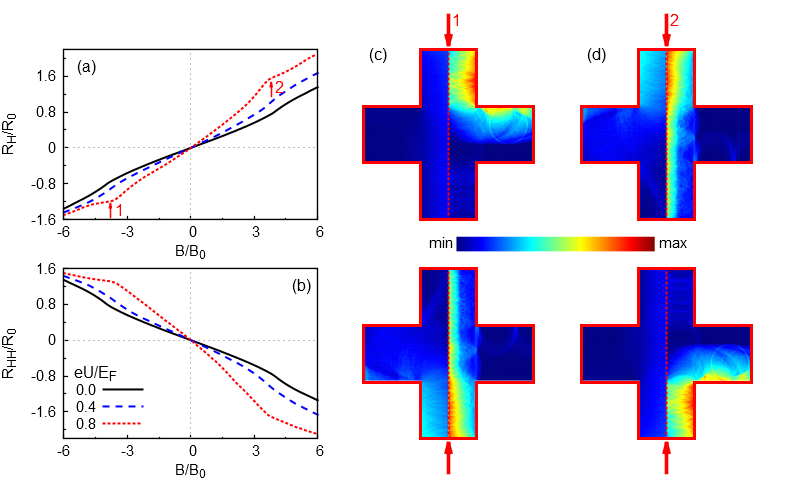}
\caption{(Color online)The Hall resistances (a) $R_H$ and (b) $R_{HH}$ for different values of applied potential $U$ given in the inset of (b). (c)-(d) Electron current density plots for magnetic field values indicated in (a). Arrows indicate the injection lead. } \label{fig2}
\end{center}
\end{figure*}
\begin{figure*}[htbp]
\begin{center}
\includegraphics[width=14cm]{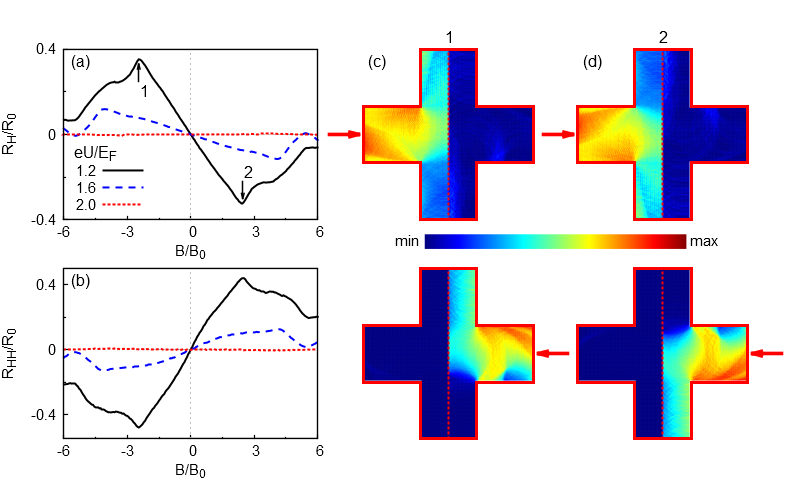}
\caption{(Color online) The same as Fig. \ref{fig2} but now for $eU>E_F$.} \label{fig3}
\end{center}
\end{figure*}
\subsection{Bilayer graphene without a gap}

In Figs. \ref{fig4} and \ref{fig4b} we plot the bend resistances as a function of the applied magnetic field. The different curves are for different values of the potential height $U$. Two distinct cases are observed: ($i$) when $0 < eU < E_{F}$ (Fig. \ref{fig4}) and ($ii$) $eU > E_{F}$ (Fig. \ref{fig4b}). Comparing with the monolayer graphene Hall bar of Ref. \onlinecite{A3} we see that the bend resistances in this system are smaller which is due to the constant DOS in
bilayer graphene. Figs. \ref{fig4}(a) and (b) show that for $eU < E_{F}$ bend resistances are none zero for a certain range of magnetic fields around $B=0$. For these values of applied potential $U$ both sides of the junction are n-type, therefore electrons that transmit through the pn-interface will preserve its direction of motion in the other region of the junction. From Fig. \ref{fig4}(a) we see that if there is no pn-junction, i.e. $U = 0$, $R_B$ is symmetric in the magnetic field as a consequence of the symmetry of the Hall device. However, introducing the pn-junction this symmetry is broken and the symmetric behavior of $R_B$ is gone as seen for $eU/E_F=0.4$. Now, the cyclotron radius on the left and the right side of the junction are different and this will affect all transmission coefficients and resistances. For $eU/E_F = 0.8$ the cyclotron radius on the right side of the junction will be more than twice smaller than the one on the left side.The effect of it is reflected in the resistances $R_B$ and $R_{BB}$, presented in Figs. \ref{fig4}(a) and (b), respectively, which are highly asymmetric. This can be also observed in the plots of the electron current density shown in Figs. \ref{fig4}(c)-(e). These plots show the flow density of carriers in the structure and represent a valuable tool for understanding the resistances. We see that at point 1, with $B=B_1$, shown in the bottom Fig. \ref{fig4}(c) electrons that are injected from lead 4 on the side of the junction having potential $U$ are unable to cross the pn-interface and the majority of them are guided to lead 2. This affects $R_B$ because $R_B\propto T_{31}T_{24} - T_{34}T_{21}$, as seen from Eq. \eqref{eq2}. If we decrease the magnetic field the cyclotron radius will increase and more electrons will be scattered to the perpendicular leads. If on the other hand the magnetic field is increased, a higher percentage of electrons is scattered back to the lead of injection - hence a peak around $B_1$. Decrease of the magnetic field leads to a higher probability that electrons will be scattered into the perpendicular leads. This will give rise to $T_{34}$ while $T_{24}$ decreases as well as the total $R_B$. At $B=0$ bend resistance drops to zero which means that the same amount of electrons is scattered to the perpendicular leads as to the opposite ones. Change of sign of $R_B$ with the change of sign of $B$ tells us that now electrons injected from lead 4 are no longer bending towards the pn-interface but towards the perpendicular leads. Similar observations can be made for points 2 and 3, with $B_2$ and $B_3$, respectively. Comparing the transmission coefficient $T_{21}$ for these two situations we can understand the presence of the decrease of the resistance around $B_3$. If we apply $B<B_2$ electrons injected from lead 1 are scattered toward all other leads, increase of $B$ beyond $B_2$ leads to smaller cyclotron radius which results in a more directional flow of electrons towards lead 4 and consequently $T_{21}\rightarrow 0$. We see that $T_{34}$ is relatively large in both cases while $T_{31}T_{24}$ is very low. Depending on the value of $T_{21}$ the bend resistance $R_B$ decreases ($0<B<B_2$) or increases towards 0 ($B_2<B<B_3$). Again, for very large field the majority of electrons are scattered back to the lead of injection and $R_B$ drops to 0.

Behavior of $R_{BB}$ can be explained in a similar way. Resistance $R_{BB}$ is obtained by switching the current and the voltage probes of the $R_B$ measurement setup. Therefore, we can say that $R_{BB} \propto T_{13}T_{42}-T_{12}T_{43}$ and the electron density plots of injection from leads 2 and 3 are needed for a detailed interpretation of the results. Fig. \ref{fig4}(b) shows similar features as Fig. \ref{fig4}(a) but for the opposite direction of magnetic field. These figures also show that as $R_B$ decreases with increase of $U$, $R_{BB}$ shows the opposite behavior which happens because now there is a different mechanism. Electrons under negative magnetic field bend towards the perpendicular leads while if a positive magnetic field is applied electrons bend towards the pn-interface which increases $T_{42}$. This is opposite to the case of $R_B$.

Figs. \ref{fig4b}(a) and (b) show the bend resistances in the case of $eU>E_F$. For these values of applied potential the right side of the junction is p-type, therefore, electrons injected from the n-region after passing the pn-interface will transform into a hole state and reverse its direction of motion. Consequently, the snake states appear along the pn-junction. Electrons reflected at the boundary will perform skipping orbits, as shown in the left part of Fig. \ref{fig1}(c), while if it is transmitted it will perform a "snake-like" movement along the pn-interface as presented in the right part of Fig. \ref{fig1}(c). This will have a dramatic impact on the resistances. As shown in the figures the bend resistances have very high negative values for magnetic field of certain sign. But if we reverse the direction of the magnetic field the behavior of the bend resistances change significantly. This can be explained using Figs. \ref{fig4b}(c)-(e) where we show electron current density plots for three values of applied magnetic field indicated in Fig. \ref{fig4b}(a) as 1, 2, 3 with $B_1, B_2, B_3$, respectively.  Fig. \ref{fig4b}(c) shows that for magnetic field $B_1$ most of the electrons injected from lead 1 will end up in lead 2, which makes $T_{21}$ large. This situation is similar to the case of Fig. \ref{fig4}(c), but here injection from lead 4 shows a different behavior. Unlike the situation presented in Fig. \ref{fig4}(c) the right side of the junction is now p-type and the holes injected in this region are moving in opposite direction than the electrons and most of them end up in lead 3 making $T_{34}$ large, while the transmission towards opposite leads is very low. For high negative magnetic fields we have $R_B\propto - T_{34}T_{21}$, and this is the reason for its high negative value. Fig. \ref{fig4b}(d) shows the situation $B_2=0$. Notice the different nature of bilayer graphene compared with monolayer graphene. Bilayer graphene doesn't exhibit the Klein tunneling phenomena, and transmission of particles through the potential step is zero for normal incidence. This can be observed in  Fig. \ref{fig4b}(d) in case of injection from lead 1. When the magnetic field is absent electrons move in straight lines and most of them arrive to the pn-interface with small incident angle and will be largely reflected back. If positive magnetic field is applied, as in Fig. \ref{fig4b}(e), we see the appearance of snake states\cite{Ma1,Ma2} in the case of injection from lead 4. These states are characteristic for the graphene pn-junctions. Electron injected in n-region along the pn-interface in the presence of the magnetic field will bend towards the pn-interface. If it passes through the p-region they change direction, and again bends towards the pn-interface and in this manner moves along it. We see that a large portion of particles injected from lead 4 forms snake states, and are guided along the pn-interface and end up in terminal 2. Now $T_{34}$ becomes very low while $T_{24}$ increases significantly and $R_B$ becomes positive. Another difference with the single layer case can be seen in plots of the bend resistances shown in Figs. \ref{fig4b}(a) and (b). For the potential step height larger than the Fermi energy we see that the peaks are shifted and they occur for nonzero magnetic field while for single layer
graphene (see Fig. 3 of Ref. \onlinecite{A3}) they were found at B = 0. Reason for this is the absence of Klein tunneling. While for monolayer graphene $T_{31}$ was largest for $B=0$, in the case of bilayer graphene we see that for $B=0$ the transmission $T_{31}\rightarrow 0$. This effect together with the snake states causes a shifting of the peaks away from zero magnetic field.

The Hall resistances $R_H$ and $R_{HH}$ shown in Figs. \ref{fig2}(a) and (b) for the potential step smaller than the Fermi energy exhibit plateau-like features, similarly as in the case of single layer graphene. These features are caused by the angle restrictions imposed by Snell's law allowing only a certain range of incident angles to transmit through the pn-interface. Compared with the single layer case   these features are less expressed here as a consequence of the square root dependence in Eq. \eqref{Angle}. Another important difference with the single layer case is that the Hall resistance is no longer antisymmetric with respect to a sign reversal of the magnetic field. This can be understood from the transmission plots shown in Figs. \ref{fig1}(c) and (d). We see that for $U/E_F \rightarrow 0$ transmission is very high for all angles, while for $U/E_F \rightarrow 1$ the range of allowed angles is smaller where the transmission is still high. Consequently, the difference between transmission from region 1 to region 2 for the opposite situation is increased. This can be clearly seen in Figs. \ref{fig2}(c) and (d). Due to this effect we have $D(B)\neq D(-B)$ and consequently $R_{H,HH}(B)\neq-R_{H,HH}(-B)$. One can use similar reasoning to find $T_{31}(B)\neq T_{13}(-B)$. Therefore, all symmetry relations found in Ref. \onlinecite{cMLG} in case of single layer graphene, e.g. $R_{B}(B) \approx R_{BB}(-B)\approx -R_{G}(-B)\approx -R_{GG}(B) $ and $R_{H}(B) = R_{HH}(-B)$, are no longer valid (except: $R_{B}(B)=-R_{G}(-B)$ and $R_{BB}(B)=-R_{GG}(-B)$). Nevertheless, Figs. \ref{fig2}  and \ref{fig3} show that the resistances still obey those symmetries approximately.

When applying a potential $eU>E_F$ in the right region causes the appearance of hole states. Change of direction of carrier movement has a strong impact on the Hall resistance. Its influence is explained in Figs. \ref{fig3}(c) and (d). Fig. \ref{fig3}(c) shows electron current density plots when the magnetic field $B=B_1$ is applied. The highest percentage of electrons injected from terminal 1 will end up in lead 2 and $T_{21}$ will be large. This transmission coefficient will give a negative contribution to $R_H$. Holes injected from lead 3 will preferably end up also in lead 2, as a consequence of the opposite direction of motion. This makes $T_{23}$ large and increases $R_H$, because $T_{23}$ gives a  positive contribution to this resistance. Therefore, in this set up injection from the n-region will decrease $R_H$ while injection from the p-region will increase it. If $B=-B_1$ is applied, as presented in Fig. \ref{fig3}(d), we see that the situation is changed. Now, injection from the left region will increase $R_H$, because the highest probability $T_{41}$ contributes positively to $R_H$ while injection from the right region decreases the resistance due to a negative contribution from $T_{43}$. Figs. \ref{fig3}(a) and (b) show that if $eU=2E_F$ is applied the Hall resistances are always zero. This was also the case with the Hall bar sculptured from a monolayer of graphene which is a consequence of the same cyclotron radius for electrons and holes in both region. The fact that electrons and holes move through the structure with the same orbiting radius but in different directions leads to $T_{23}T_{41}=T_{21}T_{43}$, and hence $R_H=0$. Same reasoning can be applied for the magnetic field dependence of $R_{HH}$.

\begin{figure}[htbp]
\begin{center}
\includegraphics[width=8cm]{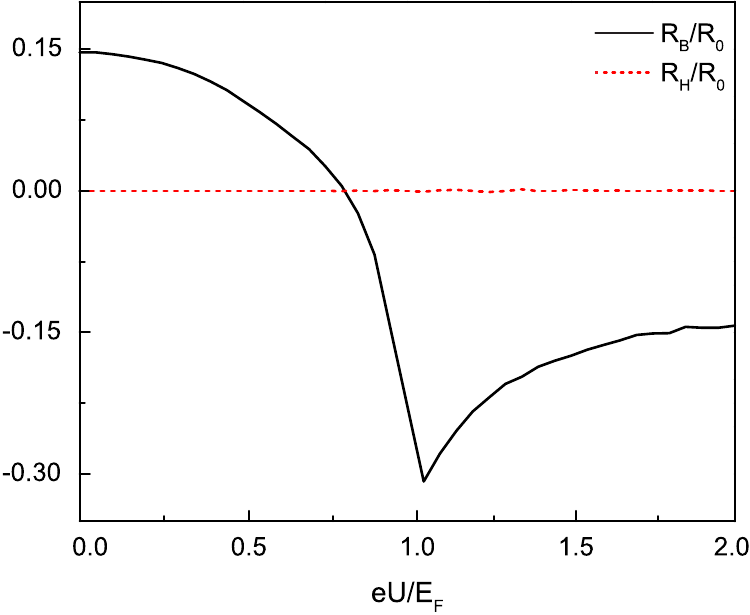}
\caption{(Color online) Bend resistance, $R_B$, and Hall resistance, $R_H$, versus potential $eU/E_F$ in the absence of a magnetic field} \label{v_array}
\end{center}
\end{figure}

Fig. \ref{v_array} shows the dependence of the bend resistance, $R_B$, and Hall resistance, $R_H$, as a function of the applied potential $U$. First noticeable feature is that $R_H$ is always zero as expected having in mind that there is no applied magnetic field. Behavior of $R_B$ can be explained in a simple way. Scattering of an electron on a potential step is expressed by Eq. \eqref{Angle} which tells us that as $|E-U_t|$ approaches zero there are less particles that can satisfy this condition and therefore will be reflected back. From the figure we can see that the dependence resembles a square root behavior as Eq. \eqref{Angle} predicts. Change of sign of $R_B$ at $eU/E_F\approx 0.8$ tells us that the measured voltage $U_{32}$ is changing sign from positive to negative which means that more electrons end up in the perpendicular than the opposite leads, as was explained for the case presented in Fig. \ref{fig4}.
\begin{figure*}[htbp]
\begin{center}
\includegraphics[width=16cm]{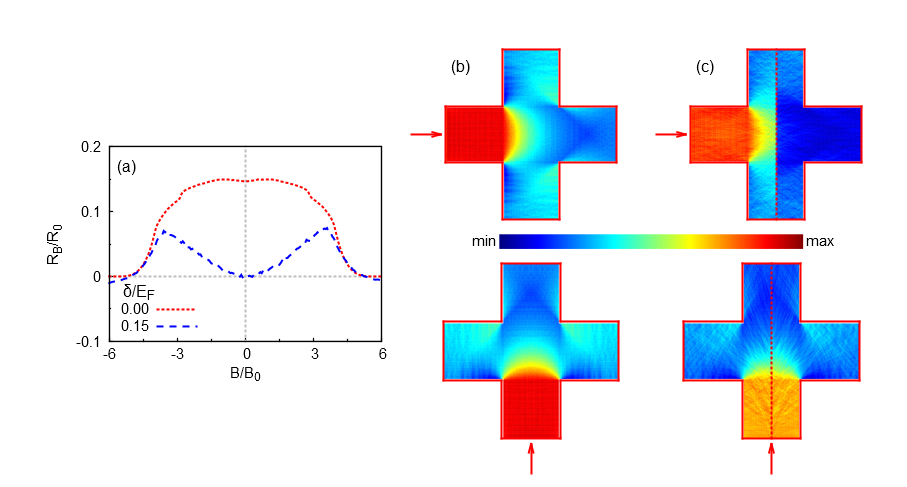}
\caption{(Color online) (a)The bend resistance $R_B$ for two cases of $\delta$ shown in the inset of (a) and $U=0$. Electron current density plots for $B = 0$ and (b) $\delta = 0$, (c) $\delta/E_F = 0.15$. Arrows indicate the injection lead. } \label{res_dvc}
\end{center}
\end{figure*}
\begin{figure*}[htbp]
\begin{center}
\includegraphics[width=16cm]{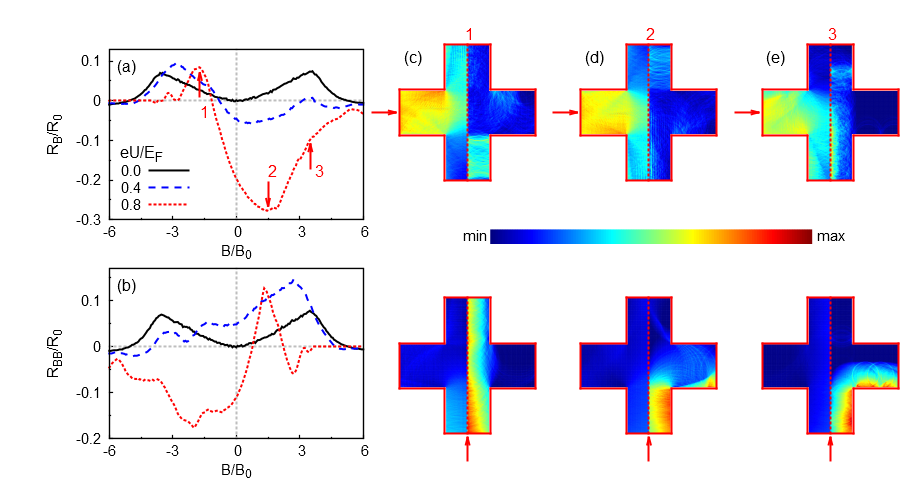}
\caption{(Color online)  The same as Fig. \ref{fig4} but now a band gap $\delta/E_F = 0.15$ is present in the second region.} \label{res_gap_1}
\end{center}
\end{figure*}
\begin{figure*}[htbp]
\begin{center}
\includegraphics[width=14cm]{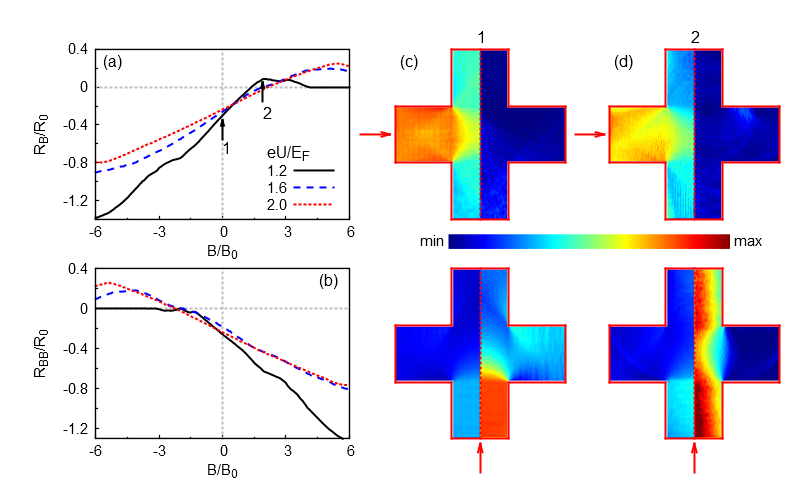}
\caption{(Color online) The same as Fig. \ref{res_gap_1} but now for $eU>E_F$.} \label{res_gap_2}
\end{center}
\end{figure*}
\subsection{Gapped bilayer graphene}
\begin{figure*}[htbp]
\begin{center}
\includegraphics[width=14cm]{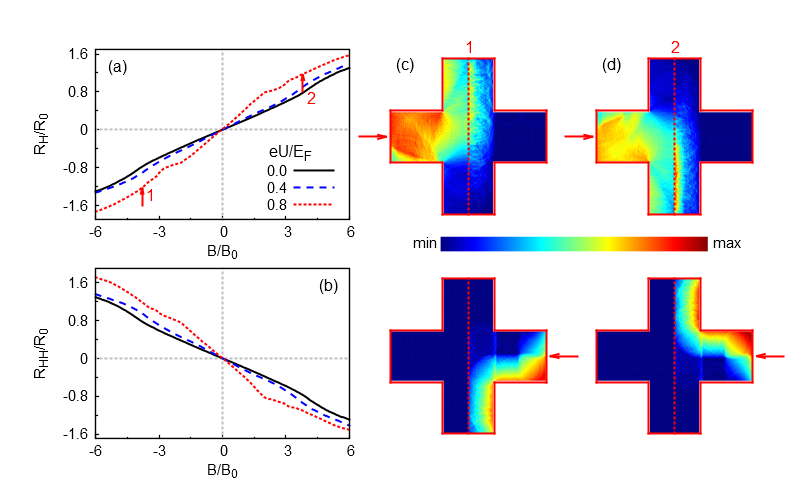}
\caption{(Color online)  The same as Fig. \ref{fig2} but now a band gap $\delta/E_F = 0.15$ is present in the second region.} \label{rhg_1}
\end{center}
\end{figure*}
\begin{figure*}[htbp]
\begin{center}
\includegraphics[width=14cm]{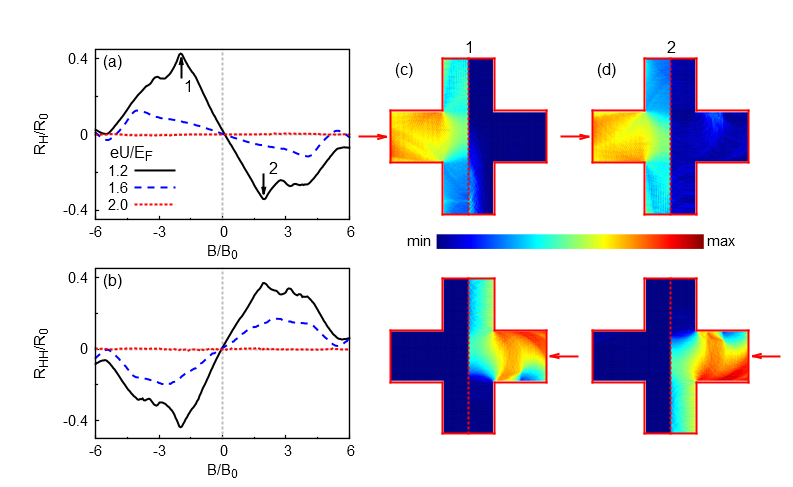}
\caption{(Color online) The same as Fig. \ref{rhg_1} but now for $eU>E_F$.} \label{rhg_2}
\end{center}
\end{figure*}

If we apply different gate potentials to the top and the bottom layer of bilayer graphene we can create a band gap in its spectrum which is tunable by the gate voltage\cite{cTBG}. In this section we will see how resistances will change if a band gap opens in the second region. Interestingly, for $U = 0$ and $B=0$ there is a minimal resistivity of the device if the band gap is opened in the second region, as shown in Fig. \ref{res_dvc}. All bend and Hall resistances drop to very small values (almost zero). Reason for this lies in the fact that for these parameters the particles that are injected from the leads perpendicular to the pn-interface (1,3) are evenly distributed over all four leads, while for the case of injection from leads parallel to the pn-interface (2,4) half of the injected particles end up in the opposite lead (4,2) while the rest is almost evenly divided between the perpendicular leads. This can be observed in Figs. \ref{res_dvc}(b) and (c).

Bend resistances for the Hall bar with only an open band gap in the second region are given in Figs. \ref{res_gap_1} and \ref{res_gap_2}. Compared with the resistances shown in Figs. \ref{fig4} and \ref{fig4b} we see that by opening a gap the resistances that suffer the most changes are the ones for which the potential $U$ lies around the Fermi energy. This is not surprising having in mind different transmission probabilities for these values of $U$. Transmissions for gapless and gapped bilayer graphene show the largest difference around $E_F$, and as we move away from it they become very similar. Still, the behavior of the resistances didn't change in a significant way. Again, all bend resistances show an asymmetric behavior for different signs of the magnetic field, for $eU>E_F$, which happens as a consequence of the snake states that appear when the hole states are present in the second region, as explained previously, while for the case $eU<E_F$ bend resistances are nonzero only for a certain range around $B=0$.

Figs. \ref{res_gap_1}(c)-(e) show electron current density plots for the same three points as in Fig. \ref{fig4}. Point 1 shows a similar peak as in the case when there is no gap in the second region, which is in agreement with the density plots shown in Fig. \ref{res_gap_1}(c). On the other hand, starting from this point we see a much faster decay of $R_B$ with magnetic field when the gap is open in the second region. Reason for this is the much smaller transmission of electrons injected from lead 1 towards the second region. Comparing Figs. \ref{fig4}(d) and \ref{res_gap_1}(d) we observe higher electron densities in the perpendicular leads when the band gap is opened in the second region. Increase of the magnetic field beyond its value at point 2 leads to a decrease of resistances, as it was for $\delta=0$. Decrease is slower in this case which can be seen at point 3, which shows higher resistance than in the configuration with no gap in the second region. This is a consequence of a much more unidirectional transport of electrons in this case. Electrons injected in lead 1 are preferably transported towards lead 4, while the electrons injected in lead 4 are moving mostly towards lead 3, as shown in Fig. \ref{res_gap_1}(e).

Bend resistances for the case when $eU>E_F$ are given in Fig. \ref{res_gap_2}. Point 1 shown in Fig. \ref{res_gap_2}(a) corresponds to zero magnetic field and has approximately the same value for $R_B$ as in the gapless graphene case. Point 2 given in the same figure shows that the peak is shifted closer to zero magnetic field when the band gap is opened. Generally this happens because the cyclotron radius becomes smaller when the band gap opens due to a shift of bands away from $U$. In order to have the same cyclotron radius as in the case of gapless graphene smaller magnetic fields need to be applied. Quantitatively, all peaks in Fig. \ref{res_gap_2} are shifted toward smaller magnetic fields but the most pronounced shifts are  for $eU\rightarrow E_F$.
\begin{figure*}[htbp]
\begin{center}
\includegraphics[width=15cm]{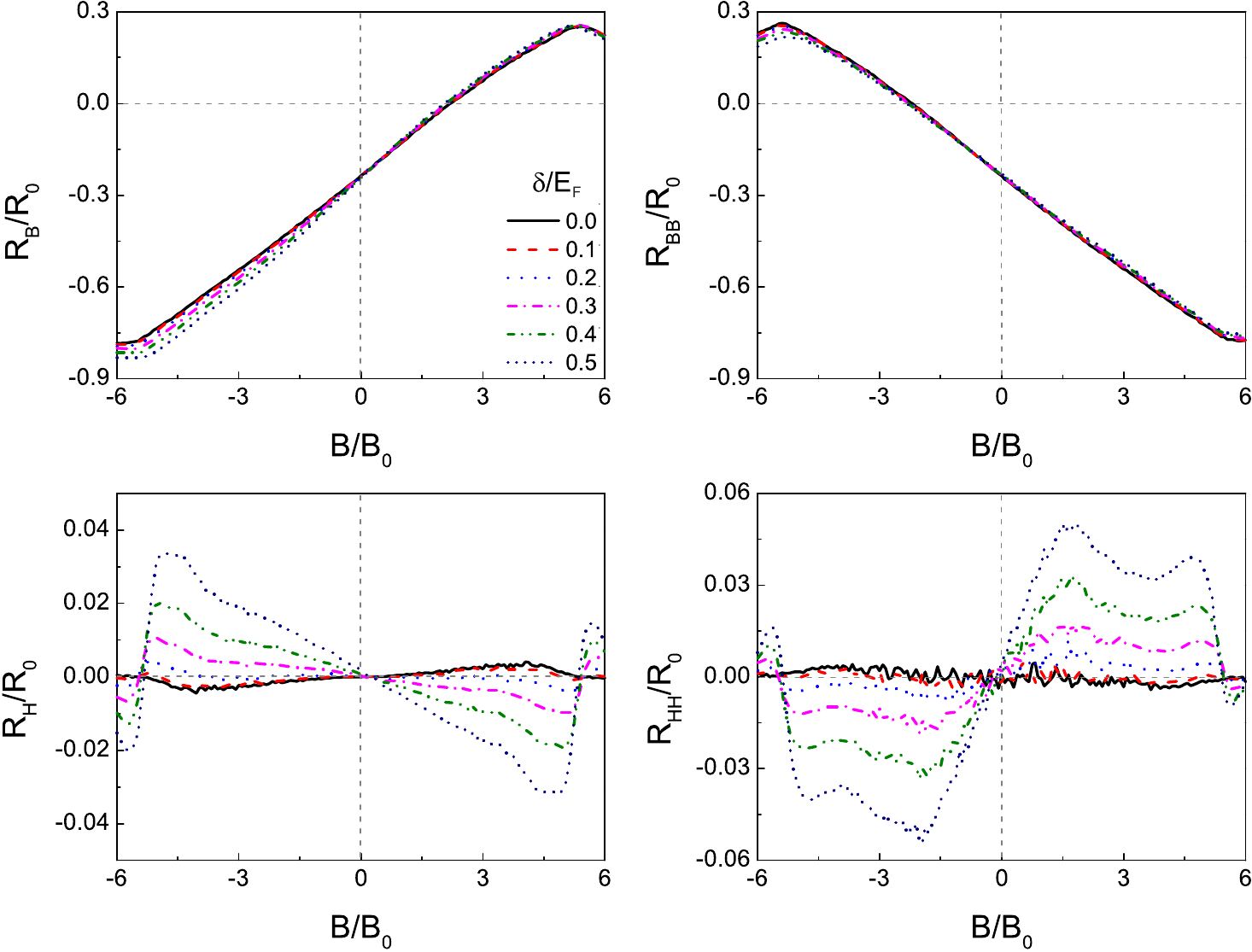}
\caption{(Color online) (a)-(b) The bend and (c)-(d) the Hall  resistances
versus magnetic field for different values of the band gap size $\delta$ shown in the inset of (a) and $eU/E_F = 2$.} \label{res_dv}
\end{center}
\end{figure*}

Hall resistances for $eU<E_F$ and $eU>E_F$ are given in Figs. \ref{rhg_1} and \ref{rhg_2}. When both sides of the junction are n-type we see an increase of the Hall resistance with magnetic field. Hall plateau-like features are almost completely absent, the only trace of it can be spotted for $eU/E_F=0.8$, when $eU\rightarrow E_F$, because only for small values of $\eps_I/ \eps_{II}= E_F/(E_F-eU)$ Snell's law will restrict the range of incident angles for which transmission is possible in a significant way so it would be visible in the plot.
Although the antisymmetric behavior of the Hall resistance is not preserved we see that the deviations from it are rather small. Reason for this can be found in Figs. \ref{rhg_1}(c) and (d) that show electron current densities for two values of magnetic field $B_2=-B_1$. Plots show very high symmetry in the case of injection from leads 1 and 3.

 If the second region is p-type the Hall resistance firstly increases with increase of the magnetic field up to some point after which it starts to decrease towards zero which happens because the carriers are moving in the opposite direction in the different regions of the junction, as was explained earlier. Plots of resistances are given in Fig. \ref{rhg_2}. Notice that the results are very similar to those obtained for gapless graphene bilayer shown in Fig. \ref{fig3} but with the major difference that the peak values are shifted towards zero magnetic field, which can be understood in the same way as for the bend resistances. Figs. \ref{rhg_2}(c) and (d) show electron current densities for the two peak values of resistance marked in Fig. \ref{rhg_2}(a). Plots show very high equivalence with the plots given in Figs. \ref{fig3}(c)-(d).

\begin{figure}[htbp]
\begin{center}
\includegraphics[width=8cm]{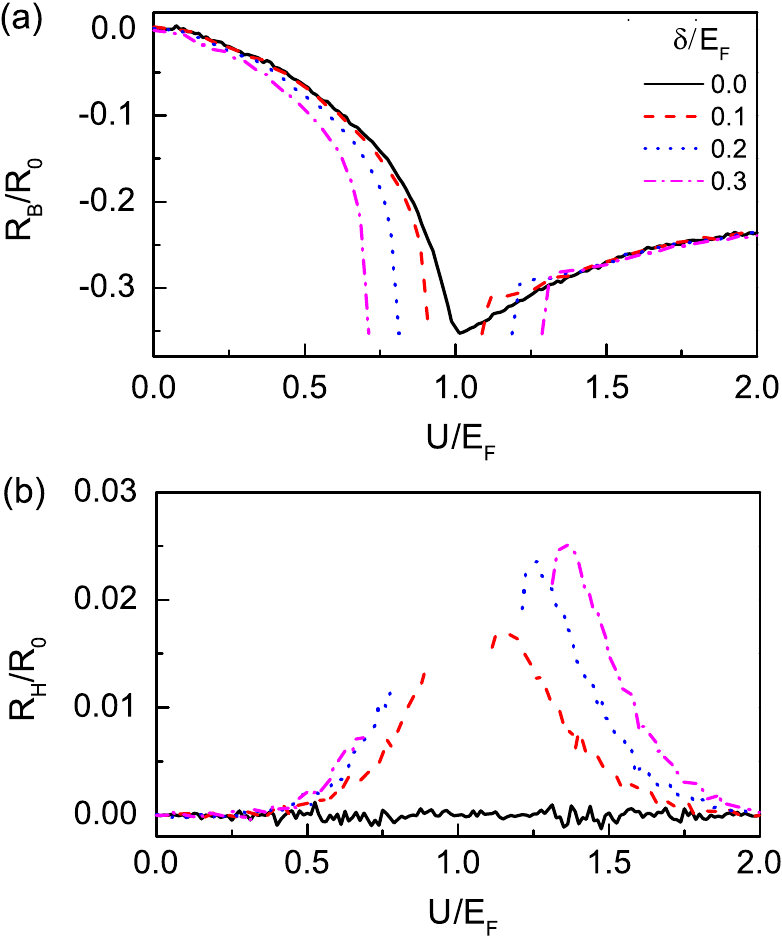}
\caption{(Color online) (a)The bend resistance $R_B$ and (b) the Hall resistance $R_H$ for different sizes of band gap shown in the inset of (a) in the case of zero magnetic field.} \label{res_dvg}
\end{center}
\end{figure}

Fig. \ref{res_dv} shows the change in resistances when we tune the band gap size. Plots are made for a system with potential step $eU/E_F = 2$. Notice that the bend resistances are not altered in a significant way when the gap size is varied between $\delta/E_F=0.1$ and $\delta/E_F=0.5$. Reason for this is the slow variation of cyclotron radius with the band gap size in case of a large potential $U$. Cyclotron radius is given by $R=\left[\sqrt{((E_F-U)^2-\delta^2)t_\perp}/(v_FqB)\right]^{1/2}$ which means that by changing the band gap from $\delta/E_F = 0.1$ to $\delta/E_F = 0.5$ the change in cyclotron radius is only $7\%$. Having in mind that the bend resistances depend quadratically on the percentage of particles that transmits to the opposite leads ($R_B\propto T_{31}T_{24}$, $R_{BB}\propto T_{13}T_{42}$, etc..) it is not surprising that they are not changed significantly with changing band gap size, especially for small magnetic fields for which the probability that the particle will end up in the opposite lead is very high. If we look at the plots of the Hall resistances given in Figs. \ref{res_dv}(c)-(d) we see that the increase of the band gap size is followed with an increase of the Hall resistance. If there is no band gap in the second region we have a pn-junction with electrons on one side rotating with a cyclotron radius $r_c$ and holes on the other side of the junction rotating with the same radius but in the opposite direction, resulting in a zero Hall resistance for all values of the magnetic field because of the symmetry in the injection from leads 1 and 3. Injection from one lead will give rise to the Hall resistance, while the injection from the other one will decrease it. Introducing a band gap in the second region breaks this symmetry which leads to a nonzero Hall resistance. Plots show that the behavior of the Hall resistances follow the one obtained in Fig. \ref{rhg_2}.

Next, we show how the bend resistance $R_B$ and the Hall resistance $R_H$ are affected by changing the potential $U$. Fig. \ref{res_dvg} shows the behavior of the resistances $R_B$ and $R_H$ for different values of $\delta$ and different applied potential in the second region. The bend resistance, given in Fig. \ref{res_dvg}(a), exhibit a steeper drop as the size of the band gap is increased. Again, we can find that the shape of the resistances follow the angle restriction imposed by Snell's law ($R_B\propto [(E_F-U)^2-\delta^2]^{1/4}$). When the charge-neutrality point (CNP) is crossed the decrease of the resistances is slowed down indicating the presence of two type of carriers moving in opposite directions preventing $R_B$ to reach zero value. The Hall resistance, on the other hand, shows an increase as the band gap is introduced. When the band gap is set to zero $R_H$ is zero due to the absence of magnetic field. If we introduce the band gap in the second region we see that $R_H$ becomes none zero which increases as $U$ approaches the Fermi energy. This happens because the pn-interface acts like a guide, electrons are unable to cross it but instead are guided to the perpendicular leads. 

Chiral states appearing inside the gap (see Fig. \ref{chrl}) can only move along the pn-interface and are therefore disconnected from the two perpendicular leads. There contribution to conduction can be measured by a to-terminal measurement. The result of such a calculation was presented in Fig. 14 of Ref. \onlinecite{cCHRL2}.
\section{Conclusion}
\label{conc}
We investigated the electronic response of a Hall bar structure sculptured from a bilayer of graphene containing a pn-junction. Simulations were done for bilayer graphene without a band gap as well as for the situation when a band gap is opened using different potentials on the top and bottom layer of the material. The Hall and bend resistances were calculated using the billiard model and the results were compared with the ones obtained for single layer Hall bars.  Although these two materials have very different properties, e.g. different energy spectrums, the absence of Veselango effect and Klein tunneling in bilayer graphene, etc., results exhibit high resemblance but with some fundamental differences. Simulations showed similarly as for the systems of single layer graphene two different transport regimes: $i$) when both regions of the junction are n-type and, $ii$) when one side of the junction is n-type while the other one is p-type. The first case showed that the Hall resistances dominate. Hall plateau-like features can be observed but unlike the single layer case they are not so strongly pronounced. Reason for this was found in the square-root energy dependence appearing in Snell's law. The second case when one of the sides of the junction is p-type showed highly asymmetric magnetic field dependence of the bend resistance. This behavior was also seen in the plots of the resistances in Refs. \onlinecite{A3} and \onlinecite{cMLG} as a consequence of the snake states emerging along the pn-interface. However, bilayer graphene shows a peak shifted away from $B=0$ which was explained with the essentially different property of these two materials, i.e. the absence of Klein tunneling in bilayer graphene (furthermore, transmission is zero for normal incidence) but the highest transmission is achieved for particular nonzero incident angles.

If we open a band gap in the second region resistances do not show dramatic changes. However, we saw that the opening of a gap almost completely removes the Hall plateau-like features in the Hall resistances for $eU<E_F$ as well as the minimal resistivity of a device for $B=0$. We also examined the dependence of the Hall resistance $R_H$ and the bend resistance $R_B$ versus the applied potential and the band gap size. Plots suggest connection in the behavior of the bend resistance with the angle restrictions imposed by Snell's law. The Hall resistance showed nonzero values for $B=0$ when the band gap is opened which occurs because the pn-interface guides the electrons to the perpendicular leads and consequently increases $R_H$.
\section{Acknowledgment}
This work was supported by the Flemish Science Foundation (FWO-Vl), the European Science Foundation (ESF) under the EUROCORES
Program EuroGRAPHENE within the project CONGRAN and the Methusalem Foundation of the Flemish government.

\end{document}